# Paradoxes of differential nonlocal cantilever beams: Reasons and a novel solution

## M. Shaat[1,2], S. Faroughi[3], L. Abasiniyan[3]

[1] *Engineering and Manufacturing Technologies Department, DACC, New Mexico State University, Las Cruces, NM 88003, USA*

[2] *Mechanical Engineering Department, Zagazig University, Zagazig 44511, Egypt*

[3] *Faculty of Mechanical Engineering, Urmia University of Technology, Urmia, Iran*

*E-mail address:*

*M. Shaat: shaat@nmsu.edu; shaatscience@yahoo.com*

*S. Faroughi: sh.farughi@uut.ac.ir*

*L. Abasiyani: l.abbasi73@yahoo.com*

## Abstract

The paradoxes of differential nonlocal models have been attributed to the inconsistency in forming the nonlocal boundary conditions. To overcome these paradoxes, nonlocal boundary conditions should be correctly formed. However, still forming these boundary conditions for differential nonlocal models is a challenging task. To resolve the trouble, in this study, the iterative nonlocal residual approach is employed. In the context of this approach, the field equation is solved in the local field with an imposed nonlocal residuals. This nonlocal residual is iteratively formed (***not determined***) depending on a pre-determined local field. The iterative nonlocal residual approach permits applying the local boundary conditions. Thus, the paradoxes of differential nonlocal models due to the nonlocal boundary conditions are effectively solved.

The paradoxes of differential nonlocal models and reasons behind them are discussed. Moreover, a nonlocal beam model based on the iterative nonlocal residual approach is proposed. This nonlocal beam model is employed to investigate the static bending and free vibration of differential nonlocal cantilever beams. Comparisons between the results of the iterative nonlocal residual approach and results of integral nonlocal beam models are carried out. It is demonstrated that the iterative nonlocal residual approach gives the same results as integral nonlocal models.

**Keywords:** Nonlocal theory; Cantilever beams; Elasticity; Paradoxes; Iterative nonlocal residual.

## 1. Introduction

Design and optimization of beam-like nano-sensors and nano-actuators for modern micro-electromechanical systems (MEMS) and nano-electromechanical systems (NEMS) gain extensive technical attentions in the current literature, see *e.g.* [1-7]. This can be attributed to the need for materials that can satisfy the small size constraint [8-11]. Moreover, these small materials provide flavors which might enhance the performance of MEMS and NEMS. For this purpose, advanced modeling techniques are proposed to replace the experiments for such small materials. For instance, both atomistic and molecular simulations have been utilized to reveal the non-traditional phenomena of small-scale materials. However, atomistic and molecular models are restricted by computational costs. Thus, utilizing them in the design and optimization of MEMS and NEMS is a challenging task. This increases the attention to the continuum mechanics-based models. These models can allow for the design and optimization of materials. However, the classical models lack the essential measures for the non-traditional phenomena of small-scale materials. Therefore, the attention in the literature is being given for developing advanced continuum mechanics models that incorporate these measures.

Nowadays, nonlocal continuum models gain an increasing consideration. The flavor that these models can give is the incorporation of measures for the non-neighbor interactions. In various small scale materials and/or systems, such nonlocal interactions are brightly observed. The foundation of nonlocal continuum mechanics can be dated back to Kröner [12], Krumhansl [13] and Kunin [14]. Later, the concept of the nonlocal elasticity has been introduced by Eringen and coworkers [15-19] for linear isotropic materials. Current nonlocal continuum models depend on Eringen's nonlocal elasticity.

Indeed, different forms were proposed for Eringen's nonlocal elasticity. First developed was the integral nonlocal model (known as fully nonlocal model). In the context of this form, an integral kernel function is employed to model the non-neighbor interactions in materials. Afterwards, Eringen [19] showed that for a Green Function-type kernel, the integral nonlocal model can be converted into an equivalent differential nonlocal model. Studies in which the differential form of Eringen's model was employed are, in fact, uncountable. For instance, the static bending, buckling, vibration, and wave propagation of nonlocal rods, beams, plates and shells were investigated using differential nonlocal elasticity [20-23]. Yu et al. [24] studied buckling of nano-beams under non-uniform temperature based on nonlocal thermo-elasticity. Barretta et al. [25] investigated Timoshenko nano-beams by employing a nonlocal Eringen-like constitutive law defined by two material length-scale parameters. More recently, Thai [26] et al. developed a simple beam theory accounting for shear deformation properties is suggested for static bending and free vibration analysis of isotropic nano-beams. The size-dependent conduct is taken by using the nonlocal differential constitutive relations of Eringen. Also, Xu et al. [27] investigated bending and buckling Euler–Bernoulli beams within the framework of the nonlocal strain gradient theory in combination with the von Kármán nonlinear geometric relation. Sari et al [28] developed a Timoshenko beam model for axially functionally graded (FG) non-uniform nano-beams with nonlocal residual based on Eringen's nonlocal theory.

In recent studies, discrepancies were revealed between the results of differential and integral nonlocal models [29-32]. For instance, it was observed that cantilever beams reflect hardening behaviors when modeled by the differential nonlocal theory and softening behaviors when modeled via the integral nonlocal theory [30]. This motivated researchers to discuss the paradoxes in the existing results of the differential nonlocal theory [30-37]. Their observations have attributed the paradoxes in differential nonlocal models to the boundary and loading conditions.

In this study, a novel solution for the paradoxes of differential nonlocal models is proposed. This solution depends on the iterative nonlocal residual approach proposed by Shaat [23]. In the context of this approach, the sum of the residual energy at a specific point due to its nonlocal interactions with the surrounding points is computed iteratively. Thus, the nonlocal residual energy is formed (*not determined*) depending on a pre-determined local field. Then, these iteratively-formed nonlocal residual energy is imposed to the local field problem. Afterwards, the field equation is solved in the local field with the imposed nonlocal residuals. This iterative nonlocal residual approach permits applying the local boundary conditions (*no need to form the nonlocal boundary conditions*), and hence the paradoxes of differential nonlocal models due to the nonlocal boundary conditions are effectively solved.

First, the paradoxes of differential nonlocal models and reasons behind them are explained. Second, a nonlocal beam model based on the iterative nonlocal residual approach is proposed. This nonlocal beam model is employed to derive solutions for differential nonlocal cantilever beams. Solutions for the static bending of nonlocal cantilever beams under point loads and distributed loads are derived. Moreover, the free vibration problem of nonlocal cantilever beams is solved. Finally, a comparison between the results as obtained utilizing the iterative nonlocal residual approach and results of integral nonlocal beam models is carried out. It is demonstrated that differential nonlocal models can give the same results as integral nonlocal models when the formal ones are solved using the iterative nonlocal residual approach.

## 2. Paradoxes in differential nonlocal beams

In fact, the paradoxes in differential nonlocal models can be observed when comparing the results of differential nonlocal beams under different boundary and loading conditions. Peddieson et al. [35] considered the static bending of differential nonlocal cantilever beams under several transverse loads. Paradoxes in the obtained results can be observed where cantilever beams subjected to point loads at the free end did not show any predictable response when changing the nonlocal parameter. Cantilever beams under uniform distributed loads, in contrast, exhibited a hardening behavior towards the nonlocal effects. Recently, the study that was performed by Peddieson et al. [35] is reconsidered by Li et al. [38]. The same results were obtained; however, it was pointed out that nanobeams may be determined with stiffer or softer stiffnesses depending on the applied loading conditions. As for the vibrational behavior of differential nonlocal beams, paradoxes can also be recognized. The fundamental frequencies of differential nonlocal cantilever beams were obtained increasing with the increase in the nonlocal parameter while the other higher-order natural frequencies were determined with decreasing trends in studies such as [39-41]. Beside these studies, nonlocal strengthening model [42-43] declares that nano-structural stiffness is enhancement with stronger nonlocal effects, while nonlocal weakening model asserts an opposite conclusion. In recent years, Li et al. [44-46] proved both two nonlocal models are correct. In other words, when increasing the nonlocal parameter, the increasing and decreasing trends of frequencies result in two kinds of differential nonlocal models, namely, the nonlocal strengthening model and nonlocal weakening model.

Paradoxes of differential nonlocal models can also be observed with comparison to integral nonlocal models. For instance, Fernández-Sáez et al. [30] introduced a general approach to solve the problem of the static bending of Euler-Bernoulli beams by employing Eringen's integral model. They showed discrepancies between the differential nonlocal beam model and the integral nonlocal beam model for various boundary conditions. Hence, they claimed that the differential nonlocal model is not an equivalent transformation to the integral nonlocal model. Khodabakhshi and Reddy [47] used a two phase nonlocal integral model and employed a finite element formulation for nonlocal beams. Their model did not fully overcome the paradoxes of cantilever beams where different static responses for consecutive values of the nonlocal parameter were obtained. Koutsoumaris et al. [48] and Eptaimeros et al. [49] developed a finite element integral nonlocal model for the static bending of beams. Two forms of the integral nonlocal model were employed and compared, namely, the two-phase nonlocal model and the fully nonlocal model. They showed the paradoxes of differential nonlocal beams when compared to their integral nonlocal model. Tuna and Kirca [31, 32] presented closed-form solutions for Euler–Bernoulli and Timoshenko beams under different loading and boundary conditions by using the integral nonlocal model. Unlike differential nonlocal models, Tuna and Kirca [31, 32] predicted consistent results for nonlocal beams with different boundary conditions.

To resolve the paradoxes associated with differential nonlocal models, some attempts were performed. For example, Challamel and Wang [33] suggested new models that combine local and nonlocal curvatures in order to resolve the paradoxes in cantilever beams. These models depend on the gradient elasticity model [50] which replaces the fully nonlocal model (integral nonlocal model) with a lower-order nonlocal model (only one neighbor interaction). To the authors' point of view, this attempt, however, did not provide a remedy for the paradoxes of the original differential nonlocal model since it presented a new nonlocal model. In another attempt, Challamel et al. [34, 51] attributed the paradoxes in differential nonlocal models to the possibility of obtaining nonself-adjoint energy functions. Therefore, to solve the paradox, they proposed a discrete microstructural model to derive the boundary conditions of cantilever beams. Although, in this attempt the boundary conditions at the free end of cantilever beams were properly derived, this approach, however, depends on a discrete microstructural model. Thus, employing this model to derive more complicated boundary conditions might be challenging, as well as it needs more analyses and efforts. In a recent study, Romano et al. [52] claimed against the majority of the existing studies on the paradoxes of differential nonlocal models. They insisted to consider the nonlocal field problem for cantilever beams does not admit solutions. However, as they mentioned, a unique solution can be derived for only one special type of boundary conditions.

The previous discussions and the aforementioned observations demonstrate only one fact that the paradoxes of differential nonlocal models are due to the inconsistency in forming the natural and essential nonlocal-type boundary conditions. Thus, as it was demonstrated in all studies on their paradoxes, differential nonlocal models can give correct solutions for nonlocal field problems under only one condition which is '*correctly forming the nonlocal boundary conditions*'. However, still forming the nonlocal-type boundary conditions for differential nonlocal models is a challenging task. Therefore, to resolve the trouble, in this study, a novel solution for the paradoxes of differential nonlocal models is proposed. This solution depends on the iterative nonlocal residual approach proposed by Shaat [23]. In the context of this approach, the local-type boundary conditions are the ones that should be employed. Thus, via this approach, no need to form the nonlocal-type boundary conditions, and hence the paradoxes are resolved.

Next, the iterative nonlocal residual approach is employed to derive consistent solutions for differential nonlocal beams. Moreover, it is demonstrated that the differential nonlocal model is an equivalent transformation for the integral nonlocal model.

## 3. Iterative nonlocal residual approach: Review

The fundamental idea of the iterative nonlocal residual approach is based up on treating iteratively-formed nonlocal residuals as imposed stress fields to a local field problem. This approach was first proposed by Shaat [23] for determining the static bending of nonlocal Kirchhoff plates. The merit of Shaat's iterative-nonlocal residual approach is that, instead of solving the field equation in the nonlocal field (*i.e. directly solving the nonlocal field equation in the nonlocal field is the origin of the existing paradoxes*), the field equation is solved in the local field where the local boundary conditions are the ones to be applied. Thus, the nonlocal residual field is formed (*not determined*) depending on a pre-determined local field. When an iterative procedure is involved, the nonlocal residual field serves as a correction of the local field problem for the nonlocal residual. Thus, with the iterations, the determined corrected-local fields converge to the nonlocal fields.

Consider a continuum body with a volume, $V$, and a surface, $S$, that is subjected to an imposed stress field, $\tau_{ij}$. Hamilton's principle can be then written in the form:

$$\int_0^{t_0} \left( -\int_V \left( \rho \frac{\partial^2 u_i}{\partial t^2} \delta u_i \right) dV + \int_V (\sigma_{ji,j} - \tau_{ji,j}) \delta u_i \, dV - \int_S n_j \sigma_{ji} \delta u_i \, dS + \int_V f_i \delta u_i \, dV \right.$$

$$\left. + \int_S \bar{\sigma}_i \delta u_i \, dS \right) dt = 0 \tag{1}$$

where $u_i$ denotes the displacement field. $f_i$ is a body force vector, and $n_j$ denotes the unit normal vector. $\rho$ is the continuum's mass density. $\sigma_{ji}$ is the local stress tensor, and $\bar{\sigma}_i$ denotes surface tractions vector. It should be noted that the essential and natural boundary conditions introduced in equation (1) are the local ones.

According to the iterative nonlocal residual approach, the imposed stress field, $\tau_{ji}$, is first formed and then imposed to the local field problem (equation (1)). This process permits the application of the local boundary conditions (*no need to form the nonlocal boundary conditions*). The imposed stress, $\tau_{ji}$, is a nonlocal-type imposed stress that is iteratively formed depending on a pre-determined local-type stress field. Therefore, for an iteration, $k$, the equation of motion can be written in the form:

$$\sigma_{ji,j}^{(k)} - \tau_{ji,j}^{(k)} + f_i = \rho \frac{\partial^2 u_i^{(k)}}{\partial t^2} \tag{2}$$

and the corresponding natural boundary conditions:

$$n_j \sigma_{ji} = \bar{\sigma}_i \tag{3}$$

In an iteration $k$, the nonlocal-type stress residual, $\tau_{ji}^{(k)}$, is formed based on the pre-determined local-type stress field in iteration $k-1$, as follows (Shaat [23]):

$$\tau_{ji}^{(k)} = \sigma_{ji}^{(k-1)} - t_{ij}^{(k-1)} \tag{4}$$

where $t_{ij}$ is the nonlocal stress. According to the differential nonlocal model, the nonlocal-type stress residual can be related to the local stress of a previous iteration as follows (Shaat [23]):

$$(1 - \xi^2 \nabla^2) \tau_{ji}^{(k)} = -\xi^2 \nabla^2 \sigma_{ji}^{(k-1)}$$

where $\sigma_{ji}^{(k-1)} = \lambda \varepsilon_{rr}^{(k-1)} \delta_{ij} + 2\mu \varepsilon_{ij}^{(k-1)}$ \hfill (5)

with $\varepsilon_{ij} = \frac{1}{2}(u_{i,j} + u_{j,i})$

where $\varepsilon_{ij}$ denotes the conventional strains. $\lambda$ and $\mu$ are the material's Lame constants. $\xi$ is the nonlocal parameter, and $\nabla$ is a 3-D gradient operator.

It should be mentioned that a detailed flow chart for the iterative procedure with an approve of convergence to the nonlocal field is found in Shaat [23].

## 4. Iterative nonlocal residual approach for Euler-Bernoulli beams

The displacement field of an Euler-Bernoulli beam can be defined as follows:

$$u_x(x,z,t) = -z\frac{\partial w(x,t)}{\partial x} \ , u_y = 0 \ , u_z(x,t) = w(x,t) \tag{6}$$

where the displacement field of a point $(x, y, z)$ belongs to the beam's centeriodal axis in the $xz$ plane, $u_i$, depends on the deflection of the beam, $w(x)$, at this point. According to the defined displacement field, the non-zero strain and local stress components can be defined as follows:

$$\varepsilon_{xx}(x,t) = -z\frac{\partial^2 w(x,t)}{\partial x^2} \tag{7}$$

$$\sigma_{xx} = -z(\lambda + 2\mu)\frac{\partial^2 w(x,t)}{\partial x^2} \ ; \sigma_{yy} = \sigma_{zz} = -z\lambda\frac{\partial^2 w(x,t)}{\partial x^2} \tag{8}$$

Moreover, Shaat [23] defined the nonlocal stress residual $\tau_{xx}$ for Euler-Bernoulli beams as follows:

$$\tau_{xx}(x,t)^{(k)} = -z(\lambda + 2\mu)\frac{\partial^2 w_c(x,t)^{(k)}}{\partial x^2} \tag{9}$$

where $w_c(x,t)$ is a deflection-correction field for the nonlocal residual.

Substituting equations (6)-(8) into equation (1), Hamilton's principle can be written for an Euler-Bernoulli beam (with a length $L$, a width $b$, and a thickness $h$) modeled according to the iterative-nonlocal residual approach as follows:

$$
\int_0^{t_0} \left( -\int_0^L \left( \rho A\frac{\partial^2 w^{(k)}}{\partial t^2}\delta w \right) dx + \int_0^L \left( \frac{\partial^2 M_{xx}^{(k)}}{\partial x^2} \right)\delta w \, dx \right.
$$
$$
+ \oint_0^L \left[ \left( \frac{\partial M_{xx}^{(k)}}{\partial x} \right)\delta w - M_{xx}^{(k)}\delta\left( \frac{\partial w}{\partial x} \right) \right] dx + \int_0^L p(x,t)\delta w \, dx \tag{10}
$$
$$
\left. -\int_0^L \left( \frac{\partial^2 \mathcal{M}_{xx}^{(k)}}{\partial x^2} \right)\delta w \, dx + \oint_0^L \left[ \overline{M}\delta\left( \frac{\partial w}{\partial x} \right) - \overline{V}\delta w \right] dx \right) dt = 0
$$

where $p(x,t)$ is an applied transverse distributed load. $\overline{M}$ and $\overline{V}$ denote, respectively, the bending moment surface traction and shear force surface traction. $M_{xx}$ denotes the local moment resultant which can be defined as follows:

$$M_{xx} = \int_A z\sigma_{xx} \, dA = -(\lambda + 2\mu)I\frac{\partial^2 w}{\partial x^2} \tag{11}$$

where $A$ and $I$ are, respectively, the beam cross-sectional area and area moment of inertia.

In equation (10), $\mathcal{M}_{xx}$ is introduced as an imposed nonlocal moment residual which can be defined as follows:

$$\mathcal{M}_{xx} = \int_A z\tau_{xx}\, dA \tag{12}$$

Consequently, the beam equation of motion in an iteration, $k$, can be written in the form:

$$\frac{\partial^2 \mathcal{M}_{xx}^{(k)}}{\partial x^2} - \frac{\partial^2 \mathcal{M}_{xx}^{(k)}}{\partial x^2} + p(x,t) = \rho A \frac{\partial^2 w}{\partial t^2}^{(k)} \tag{13}$$

and the corresponding boundary conditions can be defined as follows:

$$M_{xx}^{(k)} = \overline{M} \text{ or } \frac{\partial w}{\partial x} = \overline{\frac{\partial w}{\partial x}}$$

$$\frac{\partial M_{xx}^{(k)}}{\partial x} = \overline{V} \text{ or } w = \overline{w} \tag{14}$$

where $\overline{w}$ and $\overline{\frac{\partial w}{\partial x}}$ are prescribed deflection and slope as essential boundary conditions.

As indicated in equation (10), the local natural boundary conditions are the ones to be applied in each iteration. This presents an advantage of the proposed iterative approach in determining solutions for nonlocal field problems over the existing conventional solutions.

Because they are blended out of the same mold, the beam deflection, $w(x,t)$, and its correction field, $w_c(x,t)$, are decomposed as follows (Shaat [23]):

$$w(x,t) = W\varphi(x)e^{i\omega t}$$

$$w_c(x,t) = W_c\varphi(x)e^{i\omega t} \tag{15}$$

where $W$ denotes the amplitude of the beam deflection, and $W_c$ is the amplitude of the beam deflection-correction field. $\omega$ is the beam's natural frequency.

Because of the fact that the equation of motion (equation (13)) is solved in the local field, the mode shape function, $\varphi(x)$, in equation (15) is the one of the classical-local Euler-Bernoulli beam. By substituting equation (15) into equations (8) and (9) then substituting the result into equation (5), the following relation between the amplitude of the deflection and the amplitude of its correction field is obtained:

$$W_c^{(k)} = \psi(x)W^{(k-1)} \tag{16}$$

where

$$\psi(x) = \frac{-\xi^2 \frac{d^2\varphi(x)}{dx^2}}{1 - \xi^2 \frac{d^2\varphi(x)}{dx^2}} \tag{17}$$

Equation (17) presents $\psi(x)$ as a correction parameter.

By substituting equations (15)-(17) into the previous equations, the equation of motion can be formed in terms of the beam deflection according to the iterative nonlocal residual approach as follows:

$$W^{(k)}\left(D\frac{\partial^4\varphi(x)}{\partial x^4} - \rho A\omega^2\varphi(x)\right)e^{i\omega t} = p(x,t) + W^{(k-1)}\left(D\frac{\partial^4\varphi(x)}{\partial x^4}\psi(x)\right)e^{i\omega t} \tag{18}$$

and the corresponding boundary conditions can be defined as follows:

$$-DW^{(k)}\frac{\partial^2\varphi(x)}{\partial x^2} = \bar{M} \text{ or } \frac{\partial w}{\partial x} = \overline{\frac{\partial w}{\partial x}}$$

$$-DW^{(k)}\frac{\partial^3\varphi(x)}{\partial x^3} = \bar{V} \text{ or } w = \bar{w}$$

(19)

where $D = (\lambda + 2\mu)I$ is the beam bending rigidity.

It is should be noted that the second term of the right hand side of equation (18) presents a fictitious force which depends on the nonlocal residual, $\tau_{xx}$. At an iteration $k$, this force component is formed depending on the amplitude of a previous iteration, $k-1$. In the first iteration, the value of the fictitious force is zero, and then its value grows up with the iterations. This fictitious force is used to correct the beam deflection amplitude, $W$, for the nonlocal field effects. For a classical beam model ($\xi^2 = 0$), the beam equation of motion (18) reduces to the one for the classical Euler-Bernoulli beam where $\psi(x) = 0$.

Next, the derived iterative nonlocal residual approach for differential nonlocal Euler-Bernoulli beams is utilized to derive solutions for cantilever beams under different natural boundary conditions.

## 5. Application to differential nonlocal cantilever beams

## 5.1. Static bending of cantilever beams

The static form of equation (18) can be written as follows:

$$W^{(k)}\left[D\frac{d^4\varphi(x)}{dx^4}\right] = P(x) + W^{(k-1)}\left[D\frac{d^4\varphi(x)}{dx^4}\psi(x)\right]$$

(20)

and the boundary conditions is defined in equation (19).

By multiplying equation (20) by $\varphi(x)$ and integrating both sides over the beam length, the beam equilibrium equation for static bending can be derived in the form:

$$K_L W^{(k)} = \int_0^L P(x)\varphi(x)\,dx + K_C W^{(k-1)}$$

(21)

where the stiffnesses, $K_L$ and $K_C$, are obtained as follows:

$$K_L = \int_0^L \left(D\frac{d^4\varphi(x)}{dx^4}\varphi(x)\right)dx$$

$$K_C = \int_0^L \left(D\frac{d^4\varphi(x)}{dx^4}\psi(x)\varphi(x)\right)dx$$

(22)

where $K_L$ is the local beam stiffness, and $K_C$ is a correction stiffness.

Equation (21) can be rewritten in the following form by introducing $R = K_C/K_L$ as a nonlocal residual-based correction factor:

$$W^{(k)} = W^{(1)} + R W^{(k-1)} \text{ with } R = K_C/K_L \tag{23}$$

where $W^{(1)}$ is the amplitude of the local beam deflection:

$$W^{(1)} = \frac{\int_0^L \big(p(x)\varphi(x)\big)dx}{K_L} \tag{24}$$

The shape function $\varphi(x)$ for a cantilever beam can be defined as follows:

$$\varphi(x) = \left( \big(\cosh(\alpha x) - \cos(\alpha x)\big) - \left( \frac{\sinh(\alpha L) - \sin(\alpha L)}{\cosh(\alpha L) + \cos(\alpha L)} \right) \big(\sinh(\alpha x) - \sin(\alpha x)\big) \right) \tag{25}$$

with

$$\alpha L = 1.8751$$

Equation (23) presents an easy way to obtain the deflection of nonlocal cantilever beams. The deflection can be obtained after a few iterations utilizing the correction factor, $R$.

## 5.2. Free vibration of cantilever beams

For the free vibration of nonlocal Euler-Bernoulli beams, equation (18) can be written as follows:

$$D\left( \frac{d^4\varphi_n(x)}{dx^4} - \frac{d^4\varphi_n(x)}{dx^4}\psi_n(x) \right) - \rho A \omega_n^2 \varphi_n(x) = 0 \tag{26}$$

where $\varphi_n(x)$ is the $n$th mode shape function corresponding to the $n$th natural frequency $\omega_n$.

The multiplication of equation (26) by $\varphi_n(x)$ and the integration of the result over the beam length give the characteristics equation for the $n$th mode in the form:

$$(K_L - K_C)_n - M\omega_n^2 = 0 \tag{27}$$

where

$$M = \rho A \int_0^L \varphi_n(x)\varphi_n(x)\,dx$$

$$K_L = D \int_0^L \frac{d^4\varphi_n(x)}{dx^4}\varphi_n(x)\,dx \tag{28}$$

$$K_C = D \int_0^L \frac{d^4\varphi_n(x)}{dx^4}\psi_n(x)\varphi_n(x)\,dx$$

Consequently, the natural frequencies of the nonlocal beam can be obtained as follows:

$$\omega_n^2 = \left(\frac{K_L - K_C}{M}\right)_n = \Omega_n^2 (1 - R_n) \text{ with } R_n = \left(\frac{K_C}{K_L}\right)_n \qquad (29)$$

where $\Omega_n$ is the $n$th mode-local natural frequency:

$$\Omega_n^2 = \left(\frac{K_L}{M}\right)_n \qquad (30)$$

The shape function $\varphi_n(x)$ for a cantilever beam can be defined as follows:

$$\varphi_n(x) = \Bigg( (\cosh(\alpha_n x) - \cos(\alpha_n x))$$
$$- \left(\frac{\sinh(\alpha_n L) - \sin(\alpha_n L)}{\cosh(\alpha_n L) + \cos(\alpha_n L)}\right) (\sinh(\alpha_n x) - \sin(\alpha_n x)) \Bigg) \qquad (31)$$

with eigenvalues for the first 4 modes:

$\alpha_n L = 1.8751, 4.694, 7.8547, 10.9955$

Equation (29) presents a one-step solution for the nonlocal natural frequencies of Euler-Bernoulli beams. With no iterations, the nonlocal frequencies can be directly obtained for the different modes of vibration utilizing the correction factors, $R_n$.

## 6. Results and discussions

In this section, the static bending and the free vibration of differential nonlocal cantilever beams under different loading conditions are investigated. Utilizing the iterative nonlocal residual approach, it is demonstrated that the differential nonlocal model can give the same results as the integral nonlocal model. Thus, we demonstrate the fact that the two models are equivalent. This conclusion comments against the claims raised by Fernández-Sáez et al. [30] about the differential nonlocal model being inequivalent to the integral nonlocal model. Moreover, we demonstrate in this section that the paradoxes associated with cantilever beams are solved via the iterative nonlocal residual approach. To this end, the results of the differential nonlocal beam model based on the iterative-nonlocal residual approach are compared to the results of integral nonlocal beam models which were proposed by Fernández-Sáez et al. [30] and Tuna and Kirca [31, 32].

In the next analyses, the following nondimensional parameters are employed:

- Nondimensional deflection: $\bar{w}(x) = w(x)(D/q_0 L^4)$ where $q_0$ is the applied load intensity.
- Nondimensional natural frequencies: $\bar{\omega}_n = \omega_n \sqrt{\rho A L^4 / D}$.

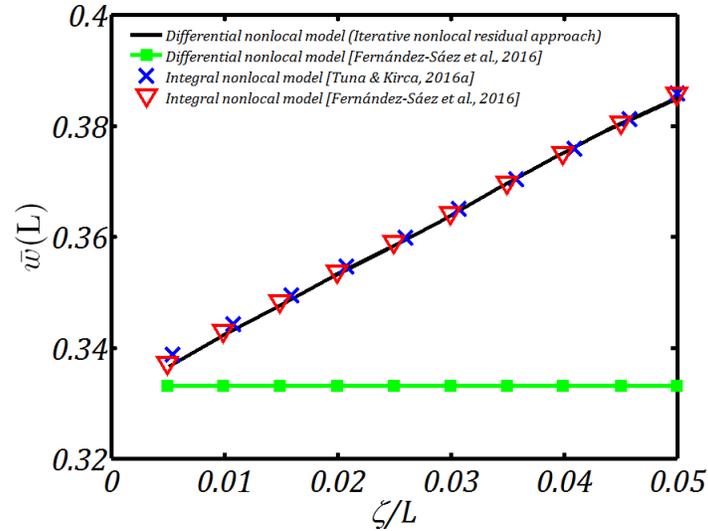

Figure 1: The free end nondimensional beam deflection, $\bar{w}(L)$, as a function of the nondimensional nonlocal parameter, $\xi/L$, for a cantilever beam subjected to a point load at its free end.

The plotted curves in Figures 1, 2, and 3 show the nondimensional maximum deflections as functions of the nondimensional nonlocal parameter ($\xi/L$) for cantilever beams under, respectively, a point load, a uniform distributed load, and a triangular distributed load. The results of the iterative nonlocal residual approach are compared to the results of conventional differential nonlocal models (Fernández-Sáez et al. 2016 [30]) and integral nonlocal models (Fernández-Sáez et al. 2016 [30]; Tuna and Kirca, 2016a [31]). These figures demonstrate the effectiveness of the iterative nonlocal residual approach to give the exact results as the integral nonlocal model. Thus, the increase in the beam deflection with the increase in the nonlocal parameter is exactly depicted as the integral nonlocal model. These results reveal that the differential nonlocal model is equivalent to the integral nonlocal model. Therefore, the claims raised by [30] that the differential nonlocal model is not an equivalent transformation for the integral nonlocal model are not correct.

Inspecting Figures 1-3, it is clear that the conventional solutions of differential nonlocal models show discrepancies when compared to solutions of integral nonlocal models or the iterative nonlocal residual approach. When differential nonlocal cantilever beams under point loads were solved using conventional methods, no nonlocal effects were revealed. In contrast, the softening mechanism of nonlocal fields on cantilever beams can be only depicted using the iterative nonlocal residual approach and integral nonlocal models. Moreover, as indicated in Figures 1 and 2, the conventional solutions for differential nonlocal cantilever beams under distributed loads reflect hardening behaviors. However, integral nonlocal models and the iterative nonlocal residual approach reflect a softening behavior for these cantilever beams. These observations indicate that the derived solutions for cantilever beams in the context of the differential nonlocal theory in previous studies, *i.e.* [35, 39-41], are incorrect. In the aforementioned studies, the moment and the shear force boundary conditions were derived depending on the differential operator. However, to be correctly formed, the natural boundary conditions should be formed depending on the integral operator.

The depicted results in Figures 1-3 demonstrate the effectiveness of the iterative nonlocal residual approach in deriving nonlocal fields with an easy way.

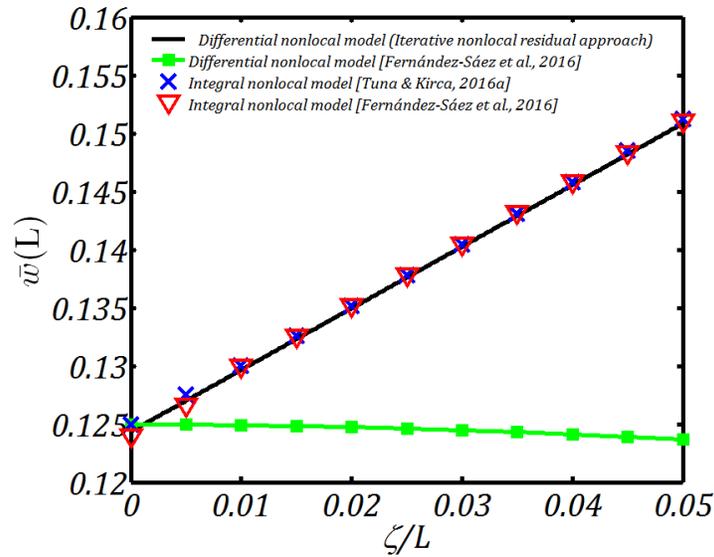

Figure 2: The free end nondimensional beam deflection, $\bar{w}(L)$, as a function of the nondimensional nonlocal parameter, $\xi/L$, for a cantilever beam subjected to a uniform distributed load.

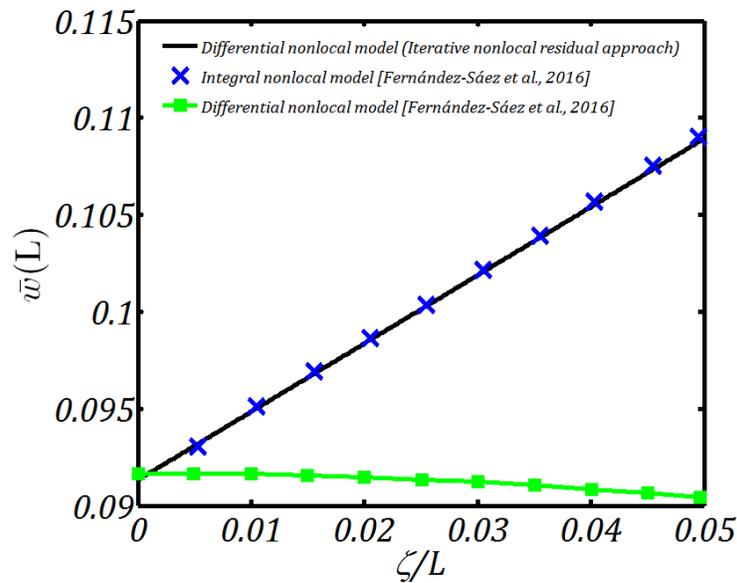

Figure 3: The free end nondimensional beam deflection, $\bar{w}(L)$, as a function of the nondimensional nonlocal parameter, $\xi/L$, for a cantilever beam subjected to a triangular distributed load ($p(x) = x$).

As for the natural frequencies of differential nonlocal cantilever beams, the first four nondimensional natural frequencies as obtained using the iterative nonlocal residual approach are compared to the results as obtained by integral nonlocal models (Tuna and Kirca [32]), gradient elasticity-based models (Challamel et al. [34]), the weighted residual approach (Xu et al. [53]), and the conventional differential nonlocal models [39, 40], as shown in Table 1.

The obtained results in Table 1 demonstrate the effectiveness of the iterative nonlocal residual approach in determining the natural frequencies of differential nonlocal beams. Conventional solutions proposed by Lu et al. [39] and Eltaher et al. [40] have shown paradoxes for cantilever beams where, unlike the high-order natural frequencies, the fundamental frequency was obtained increasing with the increase in the nonlocal parameter. These paradoxes are resolved by the iterative nonlocal residual approach. Thus, the increase in the nonlocal parameter is accompanied with reductions in the natural frequencies of all modes. The paradoxes of the conventional solutions were discussed in some studies (*e.g.* [34, 53]) where models were proposed to give the softening mechanism of cantilever beams. Although these modified models reflected the same trend of integral nonlocal models [32], discrepancies are still exist. Solutions obtained using the iterative nonlocal residual approach, on the contrary, are obtained with an excellent agreement with the solution of the integral nonlocal models proposed by Tuna and Kirca [32].

Table 1: Nondimensional natural frequencies, $\bar{\omega}_n = \omega_n \sqrt{\rho A L^4 / D}$, of cantilever beams.

| | $\xi$ | Iterative nonlocal residual approach | Tuna and Kirca [32] | Lue t al. [39] | Eltaher et al. [40] | Challamel et al. [34] | Xu et al. [53] |
|---|---|---|---|---|---|---|---|
| $\bar{\omega}_1$ | 0 | 3.51602 | 3.51602 | 3.516 | 3.5161 | 3.5160 | 3.5160 |
| | 0.05 | 3.22094 | 3.1927 | ---- | ---- | 3.4922 | ---- |
| | 0.1 | 2.92379 | 2.9187 | 3.5314 | 3.5314 | 3.4337 | 3.4369 |
| | 0.15 | 2.57183 | ---- | ---- | ---- | ---- | ---- |
| | 0.2 | 2.15599 | 2.4836 | 3.5793 | 3.547 | 3.2225 | 3.226 |
| $\bar{\omega}_2$ | 0 | 22.03364 | 22.0345 | 22.0346 | 22.0375 | 22.0345 | 22.0346 |
| | 0.05 | 18.59970 | 19.7218 | ---- | ---- | 21.1582 | ---- |
| | 0.1 | 15.85857 | 17.455 | 20.6798 | 20.6817 | 19.0866 | 19.1363 |
| | 0.15 | 13.99469 | ---- | ---- | ---- | ----- | ---- |
| | 0.2 | 12.21183 | 13.7182 | 17.5762 | 19.5111 | 14.5292 | 14.5756 |
| $\bar{\omega}_3$ | 0 | 61.69631 | 61.6972 | 61.6979 | 61.7171 | 61.6972 | 61.6979 |
| | 0.05 | 50.14570 | 53.4679 | ---- | ---- | 56.5309 | ---- |
| | 0.1 | 40.99584 | 44.3657 | 51.064 | 51.0695 | 46.5418 | 46.4933 |
| | 0.15 | 34.15174 | ---- | ---- | ---- | ---- | ---- |
| | 0.2 | 28.33317 | 30.8548 | 36.8133 | 44.5603 | 31.4454 | 31.3813 |
| $\bar{\omega}_4$ | 0 | 120.90102 | 120.9019 | 120.901 | 120.9744 | ---- | 120.9032 |
| | 0.05 | 94.71856 | 99.7959 | ---- | ---- | ---- | ---- |
| | 0.1 | 71.36360 | 76.5532 | 85.6902 | 85.6894 | ---- | 78.2659 |
| | 0.15 | 56.24601 | ---- | ---- | ---- | ---- | ---- |
| | 0.2 | 44.84016 | 48.0784 | 54.1946 | 70.0139 | ---- | 48.2693 |

## Conclusions

The paradoxes of differential nonlocal models and reasons behind them were discussed. These paradoxes were attributed to the inconsistency in forming the natural and essential nonlocal-type boundary conditions. Because forming the nonlocal-type boundary conditions for differential nonlocal models is a challenging task, the iterative nonlocal residual approach [Shaat, 2015] was utilized to propose a novel solution for the paradoxes of differential nonlocal models. It was shown that, in the framework of the iterative nonlocal residual approach, the nonlocal residual energy is iteratively formed (*not determined*) depending on a pre-determined local field. This approach permits applying the local boundary conditions where no need to form the nonlocal boundary conditions. Thus, the paradoxes of differential nonlocal models due to the nonlocal boundary conditions were effectively solved.

The iterative nonlocal residual approach was employed to solve differential nonlocal cantilever beams. Consistent results for the static deflection of cantilever beams under point loads and distributed loads were obtained. The deflection was obtained increasing with the increase in the nonlocal parameter. As for the natural frequencies, the paradoxes of the fundamental frequency was resolved. The fundamental frequency was obtained decreasing with the increase in the nonlocal parameter the same way as the other high-order mode frequencies. Furthermore, it was revealed that the differential nonlocal model solved using the iterative nonlocal residual approach gives the same results as the integral nonlocal model. The presented study demonstrates the fact that the differential nonlocal model is an equivalent transformation for the integral nonlocal model.